\documentstyle[psfig,amssym]{l-aa}
\begin{document}
 
\thesaurus{12 (11.03.1; 12.03.3; 12.03.4; 12.07.1; 12.12.1; 13.18.1)}
 
\title{Cosmology with Galaxy Clusters} 
\subtitle{III. Gravitationally Lensed Arc Statistics as a Cosmological Probe}
\author{Asantha R. Cooray}
\institute{Department of Astronomy and Astrophysics, University of Chicago, 
Chicago IL 60637, USA. E-mail: asante@hyde.uchicago.edu}
\date{Received:  July 2, 1998; accepted: September 25, 1998}
\maketitle

\begin{abstract} 
We calculate the expected number of gravitationally lensed optical, radio 
and sub-mm lensed sources
on the whole sky due to foreground galaxy clusters for different 
cosmological models. We improve previous calculations of lensed
arc statistics by including redshift information for background sources
and accounting for the redshift evolution of the foreground lensing
clusters. The background sources are described based on the 
redshift and optical magnitude or flux 
 distribution for sources in the Hubble Deep Field (HDF).
Using the HDF luminosity function, we also account for the magnification
bias in magnitude-limited observational programs to find lensed optical arcs.
The foreground lensing clusters are modeled as singular isothermal spheres,
and their number density and redshift distribution is calculated
based on the Press-Schechter theory with normalizations based on the
local cluster temperature function. 

Based on the results from optical arc surveys, we find that the 
observed number of arcs can easily be explained in
a flat universe ($\Omega_m+\Omega_\Lambda=1$) with low values 
for cosmological mass density of the universe ($\Omega_m \lesssim 0.5$).
However, given the large systematic and statistical uncertainties involved
with both the observed and predicted number of lensed arcs, more reliable
estimates of the cosmological parameters are not currently possible.
We comment on the possibility of obtaining a much tighter constraint
based on statistics from large area optical surveys. 
At radio wavelengths (1.4 GHz), 
we predict $\sim$ 1500 lensed radio sources with flux densities greater than
 $10$ $\mu$Jy, and with amplifications due to lensing greater 
than 2, in a flat cosmology with $\Omega_m=0.3$ and $\Omega_\Lambda =0.7$.
Given the recent detection of a sub-mm selected lensed  $\mu$Jy radio 
source towards A370, it is suggested
that deep radio observations of  clusters should contain such lensed
sources. At sub-mm wavelengths (850 $\mu$m), 
the number of lensed sources expected 
towards the same foreground lens population and
cosmology is $\sim 3 \times 10^{4}$. 
We briefly consider the possibility of using the South Pole 10-m sub-mm 
telescope and the Planck surveyor to identify lensed sub-mm sources.
A catalog of around 100 gravitationally lensed sources 
at 353 GHz may be a useful by-product of Planck.

\end{abstract}


\section{Introduction}

It is now well known that gravitational lensing statistics is a useful
probe of the geometry of the universe, especially for the determination
of the cosmological constant.  In a recent paper (Cooray et al. 1998a; 
hereafter CQM), we calculated the expected number
of gravitationally lensed sources in the Hubble Deep Field
(HDF; Williams et al. 1996) due to foreground galaxies as a function of
the cosmological parameters, and estimated these parameters
based on the observed lensing rate in the HDF. The expected
lensing rate was calculated based on the redshift distribution
of HDF galaxies as determined by the photometric redshift
catalogs. Similar to multiple lensing events due to foreground
galaxies, clusters of galaxies lens background sources. Such lensed
sources with high magnification appear as arcs, and the number statistics
of gravitationally lensed arcs can be used to determine the cosmological parameters
(e.g., Wu \& Mao 1996; Bartelmann et al. 1998) and study the galaxy evolution 
at high redshifts (e.g., B\'ezecourt et al. 1998).

The number statistics of lensed optical arcs have been studied by 
Wu \& Mao (1996), where they considered the effect of $\Omega_\Lambda$ on 
the predicted lensing rate, and by Bartelmann et al. (1998),
where simulations of galaxy clusters were used to calculate the
number of lensed sources. The former study relied on the spherical 
singular isothermal potential to describe foreground lensing clusters, while
the latter study used the cluster potentials observed with
numerical simulations. In between these two studies,
Hamana \& Futamase  (1997) showed that the evolution of background
source population can affect the lensing rate, while Hattori et al. 
(1997) refined the observed lensing rate by
including observational effects, such as seeing. 

In the present paper, we extend our previous
 work on the HDF (CQM) to estimate the number of lensed
optical, radio and sub-mm lensed sources
 on the sky due to foreground galaxy clusters. 
We describe
the background galaxies by the redshift and magnitude or flux
distribution of sources in the HDF. We also
assume that the HDF luminosity function, as determined
by Sawicki et al. (1997), is a valid description of the
distant universe. Thus, one of the main differences
between the present paper and previous studies involving  arc statistics
is that we use individual redshifts to calculate 
lensing probabilities, and use magnitude information to account for
various systematic effects, especially magnification bias
present in magnitude-limited optical search programs to find
lensed arcs towards galaxy clusters. A main
difference between CQM and the present work is
that we now describe the number density of
foreground lensing objects, galaxy clusters, and their evolution
using the Press-Schechter
theory (PS; Press \& Schechter 1974), normalized to the local cluster
abundance.

Similar to optical arcs, galaxy
clusters are also expected to lens background radio sources. Such lensed
sources with high magnification should appear as arcs in
radio surveys. The number statistics
of lensed radio sources can be used to determine the cosmological 
parameters, to study the radio source 
evolution at high redshifts, and as discussed later,
properties of star forming galaxies at moderate to high redshifts.
The number statistics of lensed radio sources due to 
foreground clusters were first
calculated by Wu \& Hammer (1993). They predicted $\sim$ 10 lensed radio 
sources on the sky
down to a flux density limit of 0.1 mJy, and $\sim$ 100 lensed
radio sources down to 10 $\mu$Jy at 2.7 GHz (Fig.~10 in Wu \& Hammer 1993).
At the source detection level of the VLA FIRST survey ($\sim$ 1 mJy; Becker
et al. 1995), there are only 
$\sim$ 2 to 3 lensed radio sources expected on the whole sky, 
and when compared to the area of the survey and its resolution, 
it is likely that there is no lensed source present. This prediction is 
compatible with
observational attempts to find lensed radio sources; Andernach et al. (1997)
searched the FIRST survey near Abell cluster cores and found no convincing
 candidates, 
and a statistical analysis of the radio positions towards clusters
showed no preferential tangential orientation, as expected from
gravitational lensing. Recently, a sub-mm selected source, SMM02399-0136,
towards cluster A370 was found to be lensed with an amplification of 2.5 (Ivison et al.
1998). The source was detected at 1.4 GHz, with a flux density of $\sim$ 525 $\mu$Jy.
This detection prompted us to calculate the expected number of lensed $\mu$Jy
sources present on the sky due to foreground clusters, and to refine the previous
predictions in Wu \& Hammer (1993). Since the predictions in Wu \& Hammer (1993) for sources 
down to mJy level are still expected to be valid, we will only concentrate on the $\mu$Jy sources
here. 

We also extend our calculation to estimate the number of
expected sub-mm sources on the whole sky due to foreground
clusters. Our calculation is prompted by recent observational
results from the new Sub-millimeter Common-User 
Bolometer Array (SCUBA; see, e.g., Cunningham et al. 1994)
on the James Clerk Maxwell Telescope, where a sample of gravitationally
lensed sub-mm sources has now been observed by Smail et al. (1997, 1998).
The gravitational lensing of background sub-mm sources due to
foreground galaxy clusters was first studied by Blain (1997), using
a model of a lensing cluster with predicted source counts  for background
sources. Blain (1997) showed that the surface and flux densities
of lensed sources exceed those of galaxies within the lensing cluster, and
their values. This behavior, primarily due to the slope of the
 source counts and the fact that the 
distance sources are intrinsically brighter
at sub-mm wavelengths, has now allowed the observation of moderate to
high redshift dusty star forming galaxies, which are
 amplified through the cluster potentials
(e.g., Ivison et al. 1998).
Even though lensing of sub-mm sources has been studied in literature,
no clear prediction has been made for the total number of sources
 lensed due to foreground clusters. Also,
past calculations have relied mostly  on models of background source
counts that were based on different evolutionary scenarios
for star forming galaxies. 

In Sect.~2 we discuss our calculation and its inputs.  
In Sect.~3 we present the expected number of optical, radio and sub-mm lensed
sources, and in Sect.~4 we outline possible systematic errors
involved in our calculational method. In Sect.~5, we 
compare our predicted number of
optical arcs in the whole sky to the observed number of arcs in
the Le F\`evre et al. (1994) cluster sample. In the same section
we discuss the possibility of large area optical arc survey,
using the Sloan Digitized Sky Survey (SDSS; Loveday \& Pier 1998) 
as an example.
In Sect.~5 discuss the possibility of detecting lensed $\mu$Jy
sources and sub-mm sources. A summary is presented in Sect.~6.
 We follow the conventions that the Hubble
constant, $H_0$, is 100\,$h$\ km~s$^{-1}$~Mpc$^{-1}$, the present mean
density in the universe in units of the closure density is $\Omega_m$,
and the present normalized cosmological constant is $\Omega_\Lambda$.
In a flat universe, $\Omega_m+\Omega_\Lambda=1$.

\section{Galaxy Clusters as Lenses}

In order to calculate the lensing rate for background galaxies
due to foreground galaxy clusters, we model the
lensing clusters as singular isothermal spheres (SIS) and use the
analytical  filled-beam approximation (see, e.g., Fukugita et al.
1992). In the case of the optical arcs,
the lensed sources towards clusters have all been imaged in magnitude
limited optical search programs.
Such observational surveys are 
affected by the so-called ``magnification bias''
(see, e.g., Kochanek 1991), in
which the number of lensed sources in the sample is 
larger than it would be in an unbiased sample,
because lensing brightens sources that would otherwise not be
detected.  Thus, any calculation involving
lensed source statistics should account for the magnification bias
and associated systematic effects.

We refer the reader to
CQM for full details of our lensing calculation involving
foreground galaxies as lensing sources.
Following CQM, if the probability for a source at redshift $z$ to 
be strongly lensed
is $p(z,\Omega_m,\Omega_\Lambda)$, as calculated
based on the filled-beam formalism, we can write the
number of lensed sources, $d\bar N$, with amplification
greater than $A_{\rm min}$ as:
\begin{eqnarray}
\bar N = & <F(z_l)> \times \sum_i p(z_i,\Omega_m,\Omega_\Lambda)\nonumber\\
&\times \int_{A_{\rm min}}^\infty A^{-1-\alpha(z_i)}e^{L_i/L^*(z_i)}
e^{-L_i/AL^*(z_i)}\nonumber \\
&\times \Theta(m) {2\over{(A-1)^3}}dA\nonumber\\
&\equiv <F(z_l)> \sum_i \tau(z_i) \;, 
\end{eqnarray}
where the integral is over all values of amplification $A$ greater
than $A_{\rm min}$, 
and $\alpha(z)$ and $L^{\star}(z)$ are 
parameters of the HDF luminosity function
at various redshifts as determined by Sawicki et al. 1997.
Here, the sum is over each of the galaxies in our sample.
The index $i$ represents each galaxy; hence, $z_i$ and $L_i$ are, 
respectively, the redshift, and the  rest-frame luminosity of the $i$th 
galaxy. The step function, $\Theta(m)$, takes into account
the limiting magnitude, $m_{\rm lim}$, of a given optical search to find
lensed arcs in the sky, such that only galaxies with lensed magnitude
 brighter than the limiting magnitude
is counted when determining the number of lensed arcs.
Since rest-frame luminosities of individual galaxies are
not known accurately due to uncertain K-corrections, as in CQM, we estimate
the average amplification bias by summing the
the expectation values of $\tau(z_i)$,
which were computed by weighting the integral in above equation 
by a normalized 
distribution of luminosities $L_i$ drawn from the Schechter function
at redshift $z_i$.

The probability of strong lensing depends on the number density and
typical mass of foreground objects, and is represented by the $<F(z_l)>$
in above Eq.\ 1, where $<F(z_l)>$
is the mean value over the lens redshift distribution, $z_l$.  
For SIS models, this factor
 can be written by the dimensionless parameter (CQM): 
\begin{equation} 
F\equiv 16\pi^3nR_0^3\left(\sigma_{\rm vel}\over{c}\right)^4\; ,
\end{equation} 
where $n$ is the number density of lensing objects, $R_0\equiv
c/H_0$, and $\sigma_{\rm vel}$ is the velocity dispersion in the SIS
scenario. In general, the 
parameter $F$ is independent of the Hubble constant, because the
observationally inferred  number density is proportional to $h^3$.  
In the present calculation, the foreground objects are galaxy clusters,
and thus $n$ represent the number density of clusters and $\sigma_{\rm vel}$
represent their velocity dispersion. Since the number
density of clusters in different cosmological models are expected to vary, 
we calculate the number density of galaxy clusters, $dn(M,z)$, between
mass range $(M,M+dM)$ using a  PS analysis (e.g., Lacey \& Cole 1993):
\begin{eqnarray}
\frac{dn(M,z)}{dM} = - \sqrt{\frac{2}{\pi}} & \frac{\bar{\rho}(z)}{M}\frac{d\sigma(M)}{dM} \frac{\delta_{c0}(z)}{\sigma^2(M)} \nonumber\\
&\times \exp{\left[\frac{-\delta_{c0}^2(z)}{2 \sigma^2(M)}\right]}\,
\end{eqnarray}
where $\bar{\rho}(z)$ is the mean background density at redshift $z$,
 $\sigma(M)$ is the variance of the fluctuation spectrum filtered
on mass scale $M$, and  $\delta_{c0}(z)$  is the
linear overdensity of a perturbation which has collapsed and virialized
at redshift $z$. 
Following Mathiesen \& Evrard (1998; Appendix
A), we can write $\delta_{c0}(z)$ as:
\begin{equation}
\delta_{c0} = \frac{3}{20}\left(12 \pi\right)^{\frac{2}{3}} (1+z)
\end{equation}
when $\Omega_m=1$, and
\begin{equation}
\delta_{c0}(z)= \frac{D(0)}{D(z)} \delta_c(z)
\end{equation}
when $\Omega_m+\Omega_\Lambda=1$ (flat) and $\Omega_\Lambda=0$ (open).
Here $D(z)$ is the linear growth factor and $\delta_c(z)$ is the
critical overdensity.  

For an open universe ($\Omega_\Lambda = 0$), $\delta_{c}(z)$ can be
written as (Lacey \& Cole 1993):
\begin{eqnarray}
\delta_c(z) & = & \frac{3}{2}D(z)\left(1+\frac{2\pi}{\sinh(\eta)-\eta}\right)
^{2/3} \\
D(0) & = & 1 + \frac{3}{x_0} + \frac{3\sqrt{1+x_0}}
{x_0^{3/2}}\ln(\sqrt{1+x_0}-\sqrt{x_0})
\end{eqnarray}
where $x_0 \equiv \Omega_0^{-1}-1$ and $\eta \equiv \cosh^{-1}(2/\Omega(z) - 1)$
.

For a flat universe with $\Omega_0+\Omega_\Lambda=1$, $\delta_{c}(z)$
was parameterized in Mathiesen \& Evrard (1998) as: 
\begin{equation}
\delta_c(z) = 1.68660[1+0.01256\log\Omega(z)],
\end{equation}
which was derived by Kitayama \& Suto (1997). The
$\delta_{c0}(z)$ for a flat universe was calculated using the 
linear growth factor found in Peebles (1980),
\begin{equation}
D(x)  =  \frac{\sqrt{(x^3+2)}}{x^{3/2}}\int_0^x x^{3/2}_1(x_1^3+2)^{-3/2}dx_1
\end{equation}
where $x = a/a_e$, and $a_e = [(1-\Omega_\Lambda)/(2\Omega_\Lambda)]^{1/3}$, the
inflection point in the scale factor.  This function was
integrated numerically to find the growth factor at redshifts $z$ and 0.

In addition to growth factors suggested by Mathiesen \& Evrard (1998), 
our calculation uses power-spectrum normalizations 
deduced by Viana \& Liddle (1996) for $\sigma_8 (\Omega_m)$  based on
cluster temperature function:
\begin{equation}
\sigma_8 = 0.60 \Omega_m^{0.36 +0.31 \Omega_m-0.28 \Omega_m^2},
\end{equation}
when $\Omega_\Lambda=0$, and
\begin{equation}
\sigma_8 = 0.60 \Omega_m^{0.59-0.16\Omega_m+0.06\Omega_m^2},
\end{equation}
when $\Omega_m+\Omega_\Lambda=1$.
We have also assumed
 a scale-free power spectrum $P(k) \propto k^n$
with $n=-1.4$ ($\alpha=\frac{n+3}{6} \sim 0.27$), which
corresponds to a power spectrum shape parameter $\Gamma$ of $\sim$ 0.25
in CDM models.

In order to calculate the parameter $<F(z_l)>$, we also require
knowledge of cluster velocity dispersion, $\sigma_{\rm vel}$, which
is the velocity dispersion of clusters in the SIS model. We assume that
the $\sigma_{\rm vel}$ is same as the measured velocity dispersion for
galaxy clusters based on observational data.
To relate $\sigma_{\rm vel}$ with cluster mass distribution,
we use the scaling relation between $\sigma_{\rm vel}$  
and cluster temperature, $T$, of the form (Girardi et al. 1996):
\begin{equation}
\sigma_{\rm vel}(T) = 10^{2.56 \pm 0.03} \times \left(\frac{T}{{\rm keV}}\right)^{0.56 \pm 0.05}, 
\end{equation}
and the relation between $T$ and cluster mass $M$ (Bartlett 1997;
see also Hjorth et al. 1998):
\begin{equation}
T(M,z) = 6.4 h^{2/3} \left(\frac{M}{10^{15}\, M_{\sun}}\right)^{\frac{2}{3}} (1+z)\, {\rm keV},
\end{equation}
to derive a relation between $\sigma_{\rm vel}$ and cluster mass $M$,
$\sigma_{\rm vel}(M)$.
Finally, we can write the interested parameter $F$ as a function of the
lens redshift, $z_l$:
\begin{equation}
F(z_l)=\frac{16 \pi^3}{c H_0^3} \int_{M_{\rm min}}^{\infty} \sigma(M')^4 \frac{dn(M',z_l)}{dM'}\, dM',
\end{equation}
and was numerically calculated, in additional to the above, by weighing 
over the redshift distribution of 
galaxy clusters derived based on the PS theory to obtain $<F(z_l)>$, the
mean value of $F(z_l)$, which is used in Eq.~1.

In order to compare the predicted number of bright arcs towards clusters
with the observed number towards a X-ray luminosity, $L$, selected
sample of galaxy clusters, we also need a relation between $M$ and $L$.
We obtain this relation based on the observed $L-T$ relation 
recently derived by Arnaud \& Evrard (1998):
\begin{equation}
L = 10^{45.06 \pm 0.03} \times 
\left(\frac{T}{6\, {\rm kev}}\right)^{2.88\pm0.15}\, 0.25 h^{-2}\, {\rm ergs\, s^{-1}},
\end{equation}
where $L$ is the X-ray luminosity in the 2-10 keV band,
 and $M-T$ relation in Eq.\ 13.
Since we will be comparing the predicted number of lensed arcs to the observed
number, we will be setting the minimum mass scale, $M_{\rm min}$, which
corresponds to the minimum luminosity, $L_{\rm min}$, of clusters in
optical search programs to find lensed arcs. The luminosity cutoff of
the EMSS cluster arc survey by Le F\`evre et al. (1994) is
 $8\times 10^{44}\, h^{-2}\, {\rm ergs\, s^{-1}}$, which is measured
in the EMSS band of 0.3 to 3.5 keV. By comparing the tabulated luminosities
of EMSS clusters in Nichol et al. (1997), Mushotzky \& Scharf
(1997), and Le F\`evre et al. (1994),
 we evaluate that this luminosity, in general, corresponds to
a luminosity of $\sim 12.8 \times 10^{44}\, h^{-2}\, {\rm ergs\, s^{-1}}$
in the 2 to 10 keV band.
This luminosity is calculated under the assumption of $\Omega_m=1$, and for
different cosmological parameters, it is expected that the
value will change as the  
luminosity distance relation is dependent on $\Omega_m$ and $\Omega_\Lambda$.
However, for clusters in the EMSS arc survey, with a mean redshift of
0.32, such variations are small compared to the statistical and systematic
uncertainties in the scaling relations used in the calculation. 

Ignoring various small changes due to the choice of cosmological model,
we use a minimum mass $M_{\rm min}$ of $8.8 \times 10^{14}\, h^{-1}\, 
M_{\sun}$, corresponding to above  $L_{\rm min}$. 
Using the numerical values for $\sigma(M)$ and $dn(M,z)$, and performing 
numerical integrations we find  $<F(z_l)>$ to range from $\sim$ 
$3.7 \times 10^{-6}$ when $\Omega_m=1$ to $\sim$ $2.8 \times 
10^{-4}$ when $\Omega_m=0.2$. 
 The error associated with $<F(z_l)>$ is rather uncertain.
For example, the quoted random uncertainty in $\sigma_8$ from
Viana \& Liddle (1996) is $^{+37\%}_{-27\%}$.
It is likely that $<F(z_l)>$ has an overall statistical
uncertainty of $\sim$ 50\%, however, as we discuss later, there could
also be systematic errors in our determination.

\section{Predicted Numbers}

\subsection{Optical Arcs}

In order to calculate the expected number of lensed arcs on the sky,
we use the photometric redshift catalog by Sawicki et al. (1997) for HDF 
galaxies. This catalog contains redshift information for
848 galaxies and is complete down to a magnitude of 27 in I-band.
However, HDF allows detection of sources down to a magnitude
limit of 28.5 in I-band, and contains 1577 sources down to
28 in I-band, excluding 43 apparent stars (Sawicki et al. 1997). 
We use this extra information and complemented the photometric
redshift catalog by equally distributing the additional number of 
optical sources 
between I-band magnitudes of 27 and  28, and between redshifts of 0 and 5.
Since these sources are not expected to be at very low redshifts, where
lensing probability is small, 
we do not expect to
have created a systematic bias in our study, other than perhaps
underestimate the lensing rate, if all these sources were in fact
at high redshifts.
Also, since these additional sources have 
very low magnitudes, at the limit of HDF,
we do not expect these sources to make a large 
contribution to the total number of
lensed arcs when the limiting magnitude of lensed search programs
are at the bright end. However, 
in order to calculate the true number of arcs at faint magnitudes,
it is essential that these sources be accounted for. 
The HDF galaxies are within
an area of 4.48 arcmin$^2$. We extrapolate the predicted number
of lenses in the HDF to the whole sky, by assuming that
HDF is an accurate description of the distant universe everywhere on the
sky.
Since HDF was carefully selected to avoid bright sources, 
it is likely that we have missed a large number of low redshift 
galaxies, but, such galaxies are not expected to contribute
to the lensing rate.

We have calculated the expected number
of gravitationally lensed  arcs in
by using equation (1) as a function of
$\Omega_m$ and $\Omega_{\Lambda}$, and using $A_{\rm min}$ of 10.
Since we are using the SIS model, the amplification is simply
equal to the ratio of length to width in observed lensing
arcs (see, e.g., Wu \& Mao 1996), allowing us an easy comparison
between observed number of arcs with length to width greater than 10
in Le F\`evre et al. (1994) survey.
In Table 1, we list the expected number of strongly lensed arcs in the sky
for different  $\Omega_m$ and $\Omega_\Lambda$ values, together
with the number of lensed sources at radio and sub-mm
wavelengths.

\begin{table*}
\caption[]{Predicted number of lensed optical, radio, and sub-mm sources
on the sky due to foreground clusters. $N_{\rm Planck}$ is the expected
 number of lensed sub-mm sources, with flux densities greater than 50 mJy
at 850 $\mu$m, towards clusters that are 
expected to be detected with Planck Surveyor (see, Sect.~5.3.1). The
horizontal lines across the two optical arc columns contain the current 
range of observed arc statistics (see, Sect.~5.1.1).}
\begin{flushleft}
\begin{tabular}{ccccccccccc}
\noalign{\smallskip}
\hline
\noalign{\smallskip}
& & 
\multicolumn{2}{c}{Optical ($A_{\rm min} \geq 10$)} & &
\multicolumn{2}{c}{Radio ($f_{\rm 1.4 \, GHz} \geq 10 \, {\rm \mu Jy}$)} & &
\multicolumn{3}{c}{Sub-mm ($f_{\rm 850 \, \mu m} \geq 2 \, {\rm mJy}$)} \\
\noalign{\smallskip}
\cline{3-4}  \cline{6-7}  \cline{9-11}
\noalign{\smallskip}
$\Omega_m$& $\Omega_\Lambda$ & $I_{\rm mag} \leq 22$ &
$I_{\rm mag} \leq 25$ & &  $A_{\rm min} \geq 2$ & $A_{\rm min} \geq 10$ & &
$A_{\rm min} \geq 2$ & $A_{\rm min} \geq 10$  & $N_{\rm Planck}$\\
\noalign{\smallskip}
\hline
\noalign{\smallskip}
\cline{3-4}
\noalign{\smallskip}
0.1 & 0.0 & 2190 & 8329 & & 1964 & 24 & & 14815 & 183 & 115 \\
\noalign{\smallskip}
\cline{3-4}
\noalign{\smallskip}
0.2 & 0.0 & 935 & 1880 & & 1188  & 15 & & 9910 & 123 & 75  \\
0.3 & 0.0 & 496 & 1880 & & 631 & 8 & & 6010 & 75 & 55 \\
0.4 & 0.0 & 267 & 1010 & & 343 & 4 &  & 3635 & 49 & 32 \\
0.5 & 0.0 & 150 & 568 &  & 194  & 3 &  & 2270 & 28 & 21 \\
0.6 & 0.0 & 87  & 329 & & 113 & 1.5 & & 1450 & 18 & 13 \\
0.7 & 0.0 & 52 & 197 & & 66 & 1 & & 790 & 10 & 9 \\
0.8 & 0.0 & 32 & 122 & & 42 & 0.5 & & 485 & 6 & 7 \\
0.9 & 0.0 & 20 & 81 & & 27 & 0.3 & & 310 &  4 & 4 \\
1.0 & 0.0 & 15 & 55 & & 17 & 0.2 & & 225 & 3 & 2\\
0.1 & 0.9 & 16462 & 60197 & & 15012 & 180 & & 270350 & 3340 &  2150  \\
\noalign{\smallskip}
\cline{3-4}
\noalign{\smallskip}
0.2 & 0.8 & 7017 & 25668 & & 6105 &  74 & & 101640 & 1255 & 785  \\
0.3 & 0.7 & 2970 & 10870 & & 1442 & 17 & & 37835 & 468 & 290  \\
0.4 & 0.6 & 1367 & 5010 & & 569 & 7 & & 15050 & 186 & 105 \\
\noalign{\smallskip}
\cline{3-4}
\noalign{\smallskip}
0.5 & 0.5 & 681 & 2488 & & 298 & 4  & & 6470 & 80 & 45 \\
0.6 & 0.4 & 353 & 1294 & & 163 & 2 & & 2975 & 37 & 25 \\
0.7 & 0.3 & 192 & 702 & & 94 & 1 & & 1440 & 18 & 15 \\
0.8 & 0.2 & 111 & 496 & & 58 & 1 & & 735 & 9 & 7 \\
0.9 & 0.1 & 67  & 245 & & 37 & 0.5 & & 390 & 5 & 4 \\
\noalign{\smallskip}
\hline
\end{tabular}
\end{flushleft}
\end{table*}

\subsection{Lensed Radio Sources}

In order to describe the background $\mu$Jy sources, we describe the
redshift and number distribution observed towards the HDF 
by Richards et al. (1998).
The main advantage in using the HDF data is the availability of redshift
information for $\mu$Jy sources. Also, HDF is one of the few areas where a deep
radio survey down to a flux limit of $\sim$ 2 $\mu$Jy at 1.4 GHz has been
 carried out.
The HDF contains 14 sources with flux densities of the order $\sim$ 6 to 
500 $\mu$Jy at
8.5 GHz, and  11 of these sources have measured spectroscopic redshifts.
We converted the 8.5 GHz flux densities to 1.4 GHz using individual
spectral indices as presented by Richards et al. (1998). For  sources
with no measured spectral indices,  we assumed an index of
0.4, the mean spectral index observed for $\mu$Jy sources (Fomalont et al. 1991;
Windhorst et al. 1993; Richards et al. 1998). For the 3 sources with no 
measured spectroscopic redshifts, we used photometric redshifts from the
catalog of Fern\'andez-Soto et al. (1998). 
We binned the redshift-number distribution in redshift steps of 0.25,
and calculated the lensing probability using filled-beam formalism. 
The predicted number are tabulated in Table 1 for 
minimum amplifications of 2 and 10 respectively, and down to a flux
limit of 10 $\mu$Jy at 1.4 GHz.

\subsection{Lensed Sub-mm Sources}

In order to describe the background sub-mm sources, we again use the
redshift and number distribution observed towards the HDF sources 
by Hughes et al. (1998). 
The HDF contains 5 sources with flux densities of the order $\sim$ 2 to 7
mJy. Hughes et al. (1998)  studied the probable redshifts of the detected
sources by considering the optical counterparts and assigning
probabilities for likely associations. 
In Table 1, we list the expected number of  lensed sources on the sky
for $A_{\rm min}=2$ and 10. We have also tabulated the expected
number of lensed sources towards clusters in Planck all sky survey
data, as described in Sect.~5.3.1.

\section{Systematic Errors}

Our  lensing rate
calculation relies on the assumption that the HDF is a reasonable sample of
the distance universe and that it can be applied to the whole sky.
In the case of optical arcs, 
we have included an additional number of faint sources to the
photometric redshift catalog, and by doing so, may have introduced a
systematic bias in our calculation. However, unless these sources are at either
low or high redshift, we do not expect such sources to make a large change
in the lensing rate. Also, there is a possibility that 
certain multiple sources, 
which we have counted as separate objects, may in fact represent
star-forming regions within individual galaxies (e.g., Colley et al. 1997). 
If this is true, we may have overestimated the number of sources by as much as $\sim$ 40\%,
and may have caused a systematic increase in the lensing rate.

Another possible systematic error is involved  with the
determination of the $<F(z_l)>$ parameter. We have used the
PS theory normalized to local cluster abundance
and relations between velocity dispersion, cluster temperature,
mass and luminosity to calculate $<F(z_l)>$. The used scaling relations,
as well as parameters in the PS function, have in some cases large
uncertainties. It is likely that
 our predicted numbers may be accurate to within
40\% to 50\%.  Other than statistical errors, there may also
be systematic uncertainties. For example, 
the $M-T$ relation may have additional
dependences on the cosmological parameters (see, e.g., Voit \& Donahue 1998),
which we have not fully considered. Since $<F(z_l)>$ was inferred based on 
PS function  normalized to observations, and
since these observables
depend on the assumed cosmology, $<F(z_l)>$ will also depend on it.  The
dependence of the inferred $<F(z_l)>$ on cosmology also depends on the scaling
relations, as well as the lower limit of the luminosity used in
the PS calculation, which varies with cosmology. As suggested earlier,
 for the most part, we can ignore such small
changes due to the choice of cosmological model in our
scaling relations and other observables; there are much larger
statistical and systematic errors in our calculation involving the
normalization  of the PS function etc. 

Even though we have used the PS theory to
account for redshift evolution, it is possible
that we have only partially accounted for
evolutionary effects. For example, 
we have not taken into account the effects of $\Omega_\Lambda$ on
cluster formation, where clusters are expected to be less
compact in a universe with $\Omega_\Lambda$ than in a
universe with $\Omega_\Lambda=0$. Thus, our analytical calculation is 
different from the numerical study of Bartelmann et al. (1997), where
 effects of $\Omega_\Lambda$ on 
cluster structures are accounted based on numerical simulations.
Our lens model is too simple to allow such effects, and by
ignoring this important fact, we have included an additional
simplification in our present analysis. Bartelmann et al.
(1997) found that the observed number of arcs can be explained in an
open universe, while with a cosmological constant the number predicted
is smaller than the observed statistics. However, we note that
the results from Bartelmann et al. (1997) may be in conflict with
present estimates of cosmological parameters based on other methods,
which suggest a flat universe with non-zero $\Omega_\Lambda$.
In comparison, we find that the number predicted in an open universe is
not enough to fully account for the observed statistics, unless
$\Omega_m \sim 0.1$. Also, we find a considerably
large number of arcs with a $\Omega_\Lambda$ dominated universe ($\Omega_\Lambda \sim 0.9$), which has been considered in the past to account for lensed
arcs statistics (e.g., Wu \& Mao 1996).

We have assumed that clusters can be described by singular
isothermal spheres. However, for high
amplification events such as arcs, substructures
within clusters are important; substructures are  also 
responsible for aspherical
potentials. Such potentials have been considered important for lensing
studies of individual clusters (e.g., B\'ezecourt 1998), where
it has been shown that the true lensing rate can be as high
as a factor of 2 from spherical potentials.
However, it is likely that such biases may only exist for
a small number of clusters, and thus,
for overall statistics of lensed arcs,
complex potentials can be ignored; a conclusion
also supported  by the numerical simulations of clusters
and lensed arcs. However, for certain clusters,
especially the ones that have been
systematically studied in detail due to the large number of arcs, 
which includes A2218, A370 and A1689, 
complex potentials are
important to model the individual arc distribution.

\section{Discussion}

\subsection{Lensed Optical Arcs}

\subsubsection{$\Omega_m$ from observations and Predictions}

According to Wu \& Mao
(1996), there are 9 arcs towards 39 clusters with $L > 8 \times 10^{44}\,
h^{-2}\, {\rm ergs\, s^{-1}}$ or roughly 0.2 to 0.3 arc per cluster
in the bright EMSS arc surveys
(Le F\`evre et al. 1994; Gioia \& Luppino 1994). The current predictions
for total number of clusters matching the criteria of 
EMSS arc survey clusters range from $\sim$ 7500 to 8000 (see, e.g.,
Bartelmann et al. 1997).
Thus we expect a total of $\sim$ 1500 to 1900 such arcs. This
estimate ignores the observational systematic effects in search
programs, including observational constraints such as finite seeing
 (see, e.g., Hattori et al. 1997). Since, the result  of such 
effects is to reduce the
observed number, after making an additional correction, 
we estimate a total number of 1500 to 2500 arcs on the sky, which
is slightly higher than the estimate made by Bartelmann et al. (1997).
We find that our
prediction is roughly in agreement with the observed number when
$\Omega_m \lesssim 0.4$ in a flat universe or
$\Omega_m \sim 0.1$ in an open universe. The range of
$\Omega_m$ values when $\Omega_m+\Omega_\Lambda=1$ is in agreement with our
previous estimate based on the strong lensing rate in the HDF
($\Omega_m-\Omega_{\Lambda} > -0.39$ 95\% C.I. CQM; also, Kochanek 1996),
estimates of cosmological parameters
based on the high redshift type Ia supernovae (Riess et al. 1998; $\Omega_m -
\Omega_\Lambda \sim -0.5 \pm 0.4$), 
and galaxy cluster baryonic fraction (Evrard 1997).

However, in order to derive tighter constraints on the cosmological 
parameters, we  need to consider both the statistical and systematic
errors in 
the present calculation, as well as the observed number of lensed arcs.
In general, we find that the predicted number in a $\Omega_m$ dominated
universe ($\Omega_m > 0.6$) cannot be used to explain
 the observed number of lensed
arcs, even when we consider the extreme errors in our calculation
and the observed statistics.

\subsubsection{Future Outlook}

We have predicted roughly 1500 to 3000 lensed arcs on the sky  with I-band
magnitudes greater than 22 towards foreground massive clusters.
In order to use arc statistics as a probe of the cosmological parameters, 
it is necessary that reliable results from a large area survey be used.
The current observed statistics on lensed arcs come from the optical
observations towards X-ray selected clusters in the EMSS sample, which
covers an area of $\sim$ 750 sq. degrees. In the near future,
the Sloan Digitized Sky Survey (SDSS) will take both imaging and spectroscopic
data over $\pi$ steradians of the sky. It is likely that the SDSS will image
most of the foreground massive clusters, similar to the ones
that we have considered here. The optical data from this survey are expected
to allow detection of sources down to the I band magnitude of 22.
The imaging data will be limited by seeing effects, which is expected
to limit the image resolutions to between 1.1 and 1.4 arcsecs.
For the purpose of finding lensed luminous arcs, the seeing effects would be
of a minor concern; the spatial extent of lensed arcs with length-to-width
ratios greater than 10 are not likely to be heavily affected by
observational effects. Based on our predictions, it is expected that
there will be roughly 375 to 750 arcs in the SDSS imaging data.
However, there are various practical limitations which will affect the
search for lensed arcs in SDSS data. Especially due to the large volume of 
data, it is unlikely that one would be able to select lensed arcs by just 
looking
at the images; specific algorithms to find lensed arcs are needed.
By testing such algorithms against simulated data, it is likely that
selection effects involved in the
arc search process  can be properly studied.
By considering such selection effects and the observed lensing rate
of luminous arcs, it may be possible in the future to obtain  reliable
estimates on the cosmological parameters based on arc statistics.

\subsection{Lensed Radio Sources}

We have predicted $\sim$ 1500 lensed $\mu$Jy sources, with $A>2$,
for a cosmology with $\Omega_m=0.3$ and $\Omega_\Lambda=0.7$.
The number with $A_{\rm min} > 4$ for the same cosmology is $\sim$ 200.
When compared with the lensing rate for optical arcs down to I-band
magnitude of 22 and amplifications greater than 10,
we predict a similar, or slightly lower, rate for the $\mu$Jy sources, 
down to a flux density limit of 10 $\mu$Jy. 

In comparison, Wu \& Hammer (1993) predicted $\sim$ 100 sources down to
10 $\mu$Jy towards clusters. They performed this calculation for
a cosmological model of $\Omega_m=1$, and using the X-ray
luminosity function of Edge et al. (1990). For the same cosmological model,
we predict $\sim$ 0.2 sources with amplifications greater than 10.
The difference between two predictions is primarily due to
the description of the background sources. We have used
redshift information, while Wu \& Hammer (1993) used the radio
luminosity function with no evolution assumption, an assumption
which may have overestimated the number of lensed sources.
There are also other differences between the two methods.
For example,
we have accounted for the galaxy cluster evolution for different
cosmological models using PS theory, where the number of available foreground
lensing clusters strongly decreases with an increase in the cosmological
mass density, $\Omega_m$. Such changes have not been accounted in the
previous calculation.

\subsubsection{Possibility of Detection}

Unlike optical surveys, radio surveys with interferometers such as the VLA and
the MERLIN are subjected to effects arising from instrumental limitations, 
primarily effects associated with
 resolution.  For example, there is a minimum and a maximum size for sources 
that can be detected and resolved 
with an interferometer. The largest angular scale to which the 
interferometer is sensitive restricts
the detection of high amplification sources, which are expected to appear as
 arcs, with  length to width ratios equal to amplification factors.
For the VLA A-array at 1.4 GHz, sources larger than $\sim$ 15$''$ are not 
likely
to be detected. Thus, observations of
radio arcs with length to width ratios greater than 10 may not easily be possible. In SIS model for gravitational lensing, most of the lensed
sources appear with amplification factors of 2 to 10. However, 
due to the convolution with synthesized beam,
ranging from $\sim$ 1$''$ to 5$''$, such sources are not likely to 
appear as arcs. Therefore,
detection of lensed sources with small amplifications 
are likely to be confused with foreground and cluster-member radio
sources, requiring a selection process to remove such confusing sources. 
Most of the confusion is likely to come from cluster member sources,
rather than the foreground sources, as there is an overabundance of
radio sources in clusters relative to random areas of the sky.
As discussed in Cooray et al. (1998b), based on cluster observations
at 28.5 GHz, this overabundance is likely to be high as factors of 5 to 7.
It is likely that this overabundance exists at low frequencies such as
1.4 GHz. However, certain cluster member sources
may easily be identified through source properties and
appearances; sources such as wide-angle tail sources are usually found
in cluster environments with dense IGM. Such an analysis may be limited to
few types of sources, and there is no direct radio property, such as the
radio spectral index or luminosity, that can be used to separate cluster
member sources from background ones. The identification
process of candidate lensed sources
 needs to consider the optical counterparts of radio sources;
a joint analysis between
radio and optical data may be required to recover the background radio
sources lensed through galaxy cluster potentials.
Additional observations, especially redshifts 
may be required to establish the lensed nature of $\mu$Jy sources selected 
towards clusters. This is contrary to optical searches, where lensed
galaxies can easily be established due to their arc-like appearances.

By considering the ratio between observed 
number of optical arcs and arclets and the ratio of surface 
density of optical to $\mu$Jy sources, we expect to find $\sim$ a total of
4 to 6 lensed $\mu$Jy sources 
down to 10 $\mu$Jy at 1.4 GHz towards A2218 and A370. 
For A370, one such source has already been 
recovered (Ivison et al. 1998), through the sub-mm observations of Smail 
et al. (1997). The VLA A-array 1.4 GHz data  (Owen \& Dwarakanath, 
in prep.), in which the source was detected allows detection of 
sources down to a flux limit of 50 $\mu$Jy beam$^{-1}$ (5 $\sigma$). A quick 
analysis of
the same archival data suggests that there is at least one more
$\mu$Jy lensed source towards A370 (Cooray et al., in prep.). 
It is likely that  deep surveys of
galaxy clusters with MERLIN and VLA will allow detection of $\mu$Jy 
radio sources with amplifications of 2 to 10.

As discussed in Richards et al. (1998; see, also, Cram et al. 1998), $\mu$Jy 
sources carry important information on the star formation rate and history.
Thus, observational searches for lensed sources are expected to
 allow detection of moderate to high redshift
star-forming galaxies. The search for such galaxies
will be aided by the amplification due to gravitational lensing, allowing
detections of faint sources, below the flux limits of regular surveys.
It is likely that a careful analysis of lensed $\mu$Jy sources will allow 
the study of star formation at moderate to high redshift galaxies. 
Also, the low redshift $\mu$Jy sources, 
associated with spiral galaxies are not likely to be found through clusters,
due to the low lensing rate. 
Based on our predictions and the detection of lensed
sources towards A370, we strongly recommend that deep radio observations of 
lensing clusters be carried out to find lensed sources and that such 
detections be followed up at other wavelengths.

\subsection{Lensed Sub-mm Sources}

We have predicted $\sim 3 \times 10^{4}$ lensed sub-mm sources
with flux densities greater than 2 mJy at 850 $\mu$m, and with 
amplifications greater than 2,
for a cosmology with $\Omega_m=0.3$ and $\Omega_\Lambda=0.7$.
The number with $A_{\rm min} > 4$ for the same cosmology is $\sim$ 3100,
while the number with $A_{\rm min} > 10$ is $\sim$ 500. 
We predict a lensing rate of
$\sim$ 4  sources per  cluster with amplifications greater than 2 down
to a flux limit of 2 mJy.

We compare our predicted number of lensed sources 
 to the observed number towards a sample
of galaxy clusters imaged with the SCUBA by Smail et al. (1997, 1998).
This sample contains 7 clusters with redshifts in the range $\sim$ 0.2
to 0.4. All of these clusters are well known lensing clusters in the
optical wavelengths. Unfortunately, this sample is incomplete either in
terms of X-ray luminosity or total mass. This incompleteness
doesn't allow us to perform a direct comparison between the
predicted and observed numbers. Out of the 7 clusters, 3 clusters have
X-ray luminosities greater than the lower limit imposed in our
calculation. Towards these three clusters, A370, A2390 \& A1835, there 
are 8 sub-mm sources, all of which may be gravitationally lensed.
This implies a total of $\sim 2 \times 10^{4}$ lensed sub-mm sources
on the whole sky.  Based on our lensing rate, we expect
$\sim$ 6 lensed sources towards
3 clusters; this exact number is strongly sensitive
to the cosmological parameters. Here, we have assumed a spatially-flat 
cosmological model with $\Omega_m=0.4$ and $\Omega_\Lambda=0.6$.
The predicted and observed numbers seem
 to be in agreement with each other
 for low $\Omega_m$ values in a flat universe
($\Omega_m+\Omega_\Lambda=1$). 

However, we cannot use the present observational
data to derive cosmological parameters for several reasons. These reasons
 include source contamination in the lensed source sample and systematic
biases in the foreground cluster sample. For example, it is likely that
the lensed source sample presented by Smail et al. (1998)
contain foreground and cluster-member sources. 
Since the foreground or cluster-member
sources are less bright than the background lensed sources, this
contamination is likely to be small (see, Blain 1997).
An additional systematic 
bias comes from the selection effects associated with the foreground 
 cluster sample. Since the observed clusters 
are  well known lensing clusters with high 
lensing rates at optical wavelengths,
 it is likely that there may be more lensed sub-mm 
sources towards these clusters than generally expected.
Therefore, it is likely that the Smail et al. (1998) sample is biased towards
a higher number of lensed sub-mm sources.

In order to constrain cosmological parameters based on statistics
of lensed sub-mm sources, results from a complete sample of
galaxy clusters, preferably from a large area survey,
 are needed. Further SCUBA observations of galaxy clusters,
perhaps the same cluster sample as the Le F\`evre et al. (1994) sample,
would be helpful in this regard. However, such a survey will
require a considerable amount of observing time, suggesting that current
instruments may not be able to obtain the necessary statistics.
However, in the near future there will be two opportunities to perform
 a large area sub-mm survey of galaxy clusters: 
the Planck Surveyor and the South Pole 10-m sub-mm telescope.

\subsubsection{Survey Opportunities}

{\it South Pole 10 m sub-mm telescope}---The planned South Pole (SP) 
10-m sub-mm
telescope\footnote{http://cfa-www.harvard.edu/aas/tenmeter/tenmeter.html} 
is expected to begin observations around year 2003 (see, Stark et al.
1998). At 850 $\mu$m, it is expected that
within $\sim$ 90 hours  a square degree area will be surveyed
down to a flux limit of 1 mJy. 
Given the resolution and flux sensitivity, it is likely that
the SP telescope would be an ideal
instrument to survey either a sample of clusters or  random areas
to obtain lensed source statistics down to few mJy.
To obtain reliable values of the cosmological parameters based on
the sub-mm lensed source statistics, a survey of 
several hundred square degrees down to 
few $\times$ 1 mJy will be needed. A more direct approach within
a reasonable amount of observing time would be
to survey a carefully selected sample of galaxy clusters,
either based on X-ray luminosity or total mass, from which
lensed source statistics can easily be derived.

{\it Planck Surveyor}---Considering the amplification
distribution for SIS lens model, and the number counts defined
by Scott \& White (1998), we find that roughly 100 lensed
sub-mm sources  may be detected with the 
Planck Surveyor towards galaxy clusters\footnote{http://astro.estec.esa.nl/Planck/; also, ESA document D/SCI(96)3.}. 
In Table 1, we list the number expected 
as a function of the cosmological parameters and
assuming that the Planck data will allow detection of sources
down to 50 mJy. However, given the
limited observational data on source counts at 850 $\mu$m, we note
that the predicted numbers may have large errors. We also note
that the Planck data will be highly confused, as the beam size
of Planck is $\sim$ few arcmins at 850 $\mu$m; even with $\sim$ 2 arcmin
physical pixels for high signal-to-noise  data, most of the sources
down to 50 mJy would be separated by only one or two pixels.
Assuming pixel sizes of the order beam size, the probability of finding
two sources with flux densities greater than 50 mJy in one Planck pixel would
 be $\sim$ 0.2 to 0.3. Thus, it is more likely that the Planck data will allow 
clear detection of sources down to $\sim$ 100 mJy, but with additional 
information, such as from other frequency channels and 
filtering techniques (see, e.g., Tegmark
\& de Oliviera-Costa 1998), it may be possible to lower this flux limit.

Also, it is likely that the lensed background sources
will contaminate the detection of 
Sunyaev-Zel'dovich (SZ) effect in galaxy clusters 
(see, Aghanim et al. 1997; Blain 1998). 
Given the source confusion and
contamination, it is likely that that Planck data would not readily 
allow an adequate determination of lensed sub-mm source statistics to 
constrain cosmological parameters.
It is more likely that the lensed sub-mm source catalog from Planck
 would be an important  tool to study the star-formation history at 
high redshifts; since lensing brightens sources, such a lensed source 
catalog will contain sub-mm sources  fainter than the current limit
predicted to be observable with Planck for unlensed sources.

\section{Summary}

Using the redshift and magnitude, or flux, distribution observed
towards the HDF to describe background sources and
Press-Schechter theory and singular isothermal sphere models 
to describe the foreground lensing clusters,
we have calculated the expected number of lensed arcs towards galaxy clusters.
We have improved previous calculations on arc statistics by
including redshift information for background galaxies and  accounting
for the redshift evolution of foreground lensing clusters in
different cosmological models. We have also accounted for the
magnification bias in magnitude-limited search programs based
on the HDF luminosity function. 
Our predicted numbers are in agreement with an extrapolation
of the observed number of arcs towards a sample of bright EMSS clusters
to the whole sky when
$\Omega_m \lesssim 0.5$ in a flat universe with $\Omega_m+\Omega_\Lambda=1$.
Given the large systematic effects involved with both the predicted
and observed number of arcs, more reliable constraints on the cosmological 
parameters are not currently  possible.

Using the redshift and flux information for HDF radio sources,
and the same lens population,
we have extended the calculation to
predict the  expected number of lensed  $\mu$Jy 
sources towards galaxy clusters.
In a cosmology with $\Omega_m=0.3$ and $\Omega_\Lambda=0.7$,
we predict $\sim$ 1500 lensed sources towards clusters with
X-ray luminosities greater than $12.8 \times 10^{44}\, h^{-2}\,
{\rm ergs\, s^{-1}}$, and  with amplifications due to lensing
greater than 2. 
We suggest that similar deep VLA observations may already contain
lensed $\mu$Jy sources and that a careful analysis may be required to
establish the lensing nature of such sources. 

At sub-mm wavelengths,
we predict $\sim 3 \times 10^{4}$ lensed sources towards clusters for same
cosmology as above.
We have compared our predicted numbers to 
the observed number of lensed sub-mm sources towards
a sample of galaxy clusters. However, various biases in this observed sample
and possible source contamination, do not allow us to constrain cosmological
parameters based on current statistics.
We have briefly studied the possibility of using the Planck surveyor and the 
South Pole 10-m telescope data to
perform this task. A catalog of $\sim$ 100 lensed sources towards
clusters is likely to be a useful by-product of Planck.

\begin{acknowledgements}
I would like to acknowledge useful discussions  and correspondences
with Jean Quashnock and Cole Miller on gravitational lensing,
John Carlstrom on galaxy clusters, and Heinz
Andernach, Andr\'e Fletcher, Frazer Owen and Ian Smail on gravitational lensing of radio sources due to 
foreground clusters and the possibility of an observational search to find such sources. I would also	
like to thank the referees, including Andrew Blain and serveral anonymous
referees, for their prompt refereeing of the
three separate papers on optical, radio, sub-mm lensed sources, as well as 
the combined version. I have greatly benefitted from their detailed comments and
valuable advice from Peter Schneider, which led to 
a significant improvement in this paper. 
This study was partially supported by
the McCormick Fellowship at the University of Chicago,
and a Grant-In-Aid of Research from the National Academy of 
Sciences, awarded through Sigma Xi, the Scientific Research Society.

\end{acknowledgements}

\end{document}